\documentclass[%
 reprint,
superscriptaddress,
 amsmath,amssymb,
 aps,
prl
]{revtex4-1}

\usepackage{lmodern}
\usepackage{graphicx}
\usepackage{dcolumn}
\usepackage{bm}
\usepackage{hyperref}
\hypersetup{linktocpage,colorlinks,citecolor={blue},pdfdisplaydoctitle=true,pdfpagemode=UseOutlines,bookmarksnumbered=true}
\usepackage{mathrsfs,dsfont}
\usepackage{color}

\makeatletter
\renewcommand{\@biblabel}[1]{[#1]\hfill}
\makeatother

\usepackage[normalem]{ulem}

\newcommand{\bra}[1]{\langle #1 |} 
\newcommand{\ket}[1]{| #1 \rangle } 
\newcommand{\upd}{\mathrm{d}}

\newcommand{\ie}[0]{\textit{i.e.} }
\newcommand{\eg}[0]{\textit{e.g.} }

\newcommand{\dens}[0]{\varrho}
\newcommand{\densn}[0]{\delta\varrho}

\newcommand{\denso}[0]{\hat{\varrho}}

\newcommand{\rhoo}[0]{\hat{\rho}}
\newcommand{\deco}[0]{\mathscr{D}}
\newcommand{\pairpot}[0]{\mathscr{V}}
\newcommand{\opprod}[0]{\circ}

\newcommand{\xb}[0]{\mathbf{x}}
\newcommand{\yb}[0]{\mathbf{y}}
\newcommand{\rb}[0]{\mathbf{r}}
\newcommand{\kb}[0]{\mathbf{k}}

\definecolor{cbl}{rgb}{0,0,1}
 
\definecolor{crd}{rgb}{1,0,0}

\begin{document}


\title{Principle of least decoherence for Newtonian semi-classical gravity}

\author{Antoine Tilloy}
\email{antoine.tilloy@mpq.mpg.de}
\affiliation{Max-Planck-Institut f\"ur Quantenoptik, Hans-Kopfermann-Stra{\ss}e 1, 85748 Garching, Germany}
\author{Lajos Di\'osi}
\email{diosi.lajos@wigner.mta.hu}
\affiliation{Wigner Research Centre for Physics, H-1525 Budapest 114. P.O.Box 49, Hungary}

\date{\today}
\begin{abstract}
Recent works have proved that semi-classical theories of gravity needed not be fundamentally inconsistent, at least in the Newtonian regime. Using the machinery of continuous measurement theory and feedback, it was shown that one could construct well behaved models of hybrid quantum-classical dynamics at the price of an imposed (non unique) decoherence structure. We introduce a principle of least decoherence (PLD) which allows to naturally single out a unique model from all the available options; up to some unspecified short distance regularization scale. Interestingly, the resulting model is found to coincide with the old --erstwhile only heuristically motivated-- proposal of Penrose and one of us for gravity-related spontaneous decoherence and collapse. Finally, this letter suggests that it is in the submillimeter behavior of gravity that new phenomena might be found.
\end{abstract}

\pacs{Valid PACS appear here}
\maketitle

Gluing gravity and quantum mechanics in a unified theory has proved to be a discouragingly difficult task. Most efforts have so far been focused on constructing a quantum theory of gravity \cite{kiefer2006} but the very nature of the gravitational force --classical or quantum-- remains unknown.  Whereas distinguishing different approaches to quantum gravity might for long remain out of experimental reach, pining down the classical or quantum character of gravity may be doable at low energy. To distinguish a quantum and classical gravitational force, one needs a candidate classical theory. In the Newtonian limit, the standard approach has been to consider the Schr\"odinger-Newton (SN) equation \cite{diosi1984,bahrami2014}, the low energy limit of the fundamentally semi-classical gravity of M\o ller and Rosenfeld \cite{moller1962,rosenfeld1963}, as a paradigmatic example. However, the latter is plagued by conceptual complications requiring one to drop, at the very least, the statistical interpretation of quantum states of matter. Interestingly, these foundational problems of SN are not related to relativity but, rather, to the difficulties in constructing consistent hybrid quantum-classical dynamics, a fact that has often been seen as a conclusive proof of the need for quantum gravity \cite{eppley1977}. Yet, SN is only one (rather naive) approach: we shall discuss below an alternative while other options \cite{derakhshani2014,derakhshani2016} are available.

The conceptual difficulties of hybrid quantum-classical coupling can be solved, without full quantization, provided fluctuations are added to the classical variables \cite{diosi1998,diosi2011}. The loss of unitarity can then be seen as the necessary price of semi-classical coexistence. For gravity in the Newtonian limit, it means adding a noise term $\delta \Phi$ to the Newtonian potential $\Phi$. These fluctuations are not to be derived from quantum theory (as is \eg the case in stochastic gravity \cite{calzetta1994,martin1999}) but are typically posited from heuristic considerations and required to be minimum so long as the hybrid classical-quantum coupling remains consistent. Historically, such considerations \cite{diosi1987,diosi1987newtonian,diosi1989minimum} were instrumental in the construction of models of gravity-related decoherence and collapse \cite{diosi1989,ghirardi1990,diosi2013,diosi2014,penrose1996,penrose1998,penrose2014}. The latter used Gaussian fluctuations with correlations of the form:
\begin{equation}\label{eq:fluctuations}
\mathds{E}\left[\delta \Phi_t(\xb) \delta\Phi_\tau (\yb)\right] = \text{const} \times \frac{G\hbar}{|\xb-\yb|} \delta(t-\tau),
\end{equation}
but with admittedly vague physical justifications.

A rigorous formulation of these early conjectures can be obtained via theories that rely, at least at a formal level, on the machinery of continuous measurement and control theory \cite{jacobs2006,wiseman2009}. The idea is to use the ``signal'' from an hypothetical measurement scheme to create a classical attractive force between particles via feedback control. By design, such a method yields consistent semi-classical dynamics with the smallest amount of classical randomness. It has been instantiated with two different objectives in mind. Kafri \textit{et al.} used it to constrain the theories of gravity that could be created from local operations and classical communications (LOCC), an important constraint on interactions between quantum particles used extensively in quantum information theory. The concept was illustrated first perturbatively in two-body interactions \cite{kafri2014}, then 
in a non-perturbative many-body system, albeit in a discretized space \cite{kafri2015}. Independently in \cite{tilloy2016}, a similar formalism was leveraged to build a parallel with models of objective (or spontaneous) wave function collapse used in the foundations of quantum theory (see \cite{bassi2013} for a review), emphasizing the important conceptual clarification such an approach could yield; in particular with the suppression of macroscopic superpositions. In this latter instance \cite{tilloy2016}, the full Newtonian gravitational interaction could be introduced directly in continuous space but the possible LOCC character of gravity was not discussed. 

Our objective is to revisit this second construction and show that it naturally suggests the introduction of a principle of least decoherence (PLD), less restrictive than the concept of LOCC, that is still arguably compatible with a ``classical'' character of gravity. Further, we want to show that this principle singles out classical fluctuations of the form \eqref{eq:fluctuations} and thus a model (DP) of gravity-related decoherence and collapse  introduced by Penrose and one of us almost 30 years ago \cite{diosi1989,ghirardi1990,penrose1996}. Finally, we discuss the PLD in a weaker form to revisit the concept of LOCC and suggest a natural relation between the two parameters of the Continuous Spontaneous Localization model (CSL) \cite{ghirardi1990csl}.

\paragraph{General framework --} 
In contrast with \cite{kafri2014,kafri2015}, we construct the classical Newton field $\Phi(\xb)$ to mediate the gravitational interaction. Although gravity looks like a pair-wise force 
in Newtonian limit of General Relativity, it seems important to have a concept of field available already in the non-relativistic setting.
The central question thus becomes to define a classical mass density field $\dens$ to source the gravitational field from quantum matter,
\begin{equation}\label{eq:poisson}
    \nabla^2 \Phi(\xb)=4\pi G \dens(\xb).
\end{equation}
Indeed, for a given gravitational field
\begin{equation}\label{eq:Nfield}
    \Phi(\xb)=\int\frac{-G}{\vert\xb-\yb\vert}\dens(\yb)\upd\yb
                     \equiv\int\pairpot(\xb,\yb)\dens(\yb)\upd\yb,
\end{equation}
with the standard Laplace Green function
\begin{equation}
\pairpot(\xb,\yb)=\left[\frac{4\pi G}{\nabla^2}\right](\xb,\yb)=-\frac{G}{\vert\xb-\yb\vert} \label{eq:pairpotential},
\end{equation}
the external potential showing up in the Schr\"odinger equation is then uncontroversial:
\begin{equation}\label{eq:potential}
    \hat{V}=\int \upd \xb \, \Phi(\xb) \denso (\xb),
\end{equation}
where $\denso$ is the mass density operator \footnote{For $n$ species of particles of mass $m_{\ell}$, the mass density operator reads:
\[\denso(\xb)=\sum_{\ell=1}^n m_\ell\, a^\dagger_\ell(\xb) a_\ell(\xb), \] where $a^\dagger_\ell(\xb)$ is the standard creation operator for the species $\ell$ in $\xb$ (obeying the usual commutation or anti-commutation relations according to the statistics of the particle). }. This potential does not, by itself, create entanglement between distant particles. As we will see, it is the generation of the field $\Phi$ which may break LOCC.

The mass density sourcing the gravitational field should be a ``tangible'' variable \cite{diosi2012}, \ie a field that can back-react on matter without destroying the statistical interpretation of the state vector. The usual choice, yielding the SN equation, is to take $\dens = \bra{\psi_t}\denso \ket{\psi_t}$, but it manifestly makes the resulting quantum state evolution containing $\hat{V}$ non-linear; quantum expectation values are not tangible in that sense. As we have argued, some additional noise is necessary. We shall obtain the suitable noise structure by considering the noisy signal (or continuous readout) one would obtain from the continuous measurement of the mass density operator, instead of its expectation value.

We recall the general framework introduced in \cite{tilloy2016}. For clarity, we postpone the discussion of the crucial issue of regularization and the following equations are thus momentarily divergent. The signal from the continuous measurement of the mass density reads:
\begin{equation}\label{eq:signaldefinition}
    \dens_t(\xb) = \langle \denso(\xb) \rangle_t + \densn_t(\xb),
\end{equation}
where $\langle\cdot \rangle_t=\bra{\psi_t} \cdot \ket{\psi_t}$ with $\ket{\psi_t}$ the many-particle system quantum state and where $\densn_t(\xb)$ is white noise in time, characterized by the correlation function:
\begin{equation}\label{eq:noisecorrelation} 
\mathds{E}\left[\densn_t(\xb) \densn_\tau(\yb)\right]=\gamma^{-1}(\xb,\yb)\, \delta(t-\tau), 
\end{equation}
where $\gamma^{-1}$ is the (positive semi-definite) precision kernel of the monitoring (the operator inverse of $\gamma$, the monitoring strength). The deviation of the kernel from a Dirac delta encodes the entanglement of the fictitious detectors at different space-time points (which generically breaks the LOCC assumption of \cite{kafri2014}). For a given monitored mass density field variable $\dens$, continuous measurement theory \cite{jacobs2006,wiseman2009} fixes the evolution of the corresponding matter pure state $\ket{\psi}$ \cite{tilloy2016}:
\begin{align}\label{eq:sse}
\frac{\upd}{\upd t}\ket{\psi_t}=&-i\hat{H}\,\ket{\psi_t} -\frac{1}{8}
\int \upd \xb \upd \yb\, \gamma(\xb,\yb)\left(\denso (\xb) - \langle
\denso(\xb) \rangle_t \right) \nonumber\\
&\times\left[ \left(\denso (\yb) - \langle \denso(\yb) \rangle_t
\right)-4\densn_t(\yb)\right]\ket{\psi_t},
\end{align}
where the multiplicative noise term is understood in the It\^o convention and henceforth
we set $\hbar=1$. One can then obtain the monitoring master equation (ME) for $\rhoo_t=\mathds{E}\big[\ket{\psi_t}\bra{\psi_t}\big]$ from \eqref{eq:sse} using It\^o's lemma:
\begin{equation}
    \frac{\upd \rhoo}{\upd t}\! = -i\big[\hat{H},\rhoo\big] -\frac{1}{8} \int  \!\upd\xb \upd\yb \,\gamma(\xb,\yb)\big[\denso(\xb),\left[\denso(\yb),\rhoo\right]\big],
\end{equation}
where we omit time indices for compactness. This equation so far does not contain gravity. Introducing carefully \cite{tilloy2016} the potential $\hat{V}$ of \eqref{eq:potential} in \eqref{eq:sse} yields a joint stochastic evolution for the quantum-classical couple $\{\ket{\psi_t},\Phi_t(\xb)\}$, where the two elements back-react on each other. The derivation of the corresponding stochastic master equation is provided in \cite{tilloy2016} and yields, after averaging over the noise, a ME that can be rewritten this way:
\begin{equation}\label{eq:fullME}
\begin{split}
    \frac{\upd \rhoo}{\upd t}\! =& -i\left[\hat{H}+\frac{1}{2}\int  \!\upd\xb \upd\yb \, \pairpot (\xb,\yb) \denso (\xb) \denso (\yb)\; ,\rhoo\right] \\
    &-\frac{1}{2} \int  \!\upd\xb \upd\yb \,\deco (\xb,\yb)\big[\denso(\xb),\left[\denso(\yb),\rhoo\right]\big],
    \end{split}
\end{equation}
with the new decoherence kernel:
\begin{equation}\label{eq:decoherencefunctional}
\deco(\xb,\yb)=\left[\frac{\gamma}{4} + \mathscr{V}\opprod\gamma^{-1}\opprod \mathscr{V}^\top\right](\xb,\yb)
\end{equation}
where ``$\opprod$'' denotes the product of two kernels: 
\begin{equation}
    (h_1\opprod h_2)(\xb,\yb)=\int \upd \rb\, h_1(\xb,\rb)\,  h_2(\rb,\yb).
\end{equation} 
Hence, in this framework, the fact that gravity is mediated by a classical field adds decoherence to the standard Newtonian pair potential. We should emphasize that in this context, continuous measurement theory is but a tool to derive consistent hybrid quantum-classical dynamics and ME \eqref{eq:fullME}, stripped, as stated in \cite{tilloy2016}, of its usual instrumentalist interpretation. The reader may thus consider that the final ME \eqref{eq:fullME} we obtain is that of a spontaneous collapse (or dynamical reduction) model \cite{bassi2013} which is taken as fundamental.

\paragraph{Principle of least decoherence --} 
This approach to Newtonian semi-classical gravity still possesses a lot of arbitrariness in the monitoring strength $\gamma$. Imposing an LOCC constraint as Kafri \textit{et al.}, this time in our classical-field-mediated gravity,
would single out a Dirac delta. We shall consider this option later. Our main objective is to introduce a principle that allows to single out another natural kernel, relaxing the requirement of LOCC.

The decoherence kernel $\deco$ \eqref{eq:decoherencefunctional} consists of two terms. The first term
$\gamma/4$ is the ``price'' we pay for obtaining the monitored signal $\dens$
\eqref{eq:signaldefinition} at precision/noise $\delta\dens$.
The second term comes from the noisiness in the gravitational back-reaction mediated by the signal \eqref{eq:potential}.
The latter is inversely proportional to the former: reducing the noise in the signal requires
a stronger monitoring which in turn leads to a larger decoherence, and vice versa. 
This suggests that the total decoherence $\deco$ may have a minimum, enforcing a \emph{principle of least decoherence} (PLD), for a special choice of kernel $\gamma$. 
Since all kernels appearing  in \eqref{eq:decoherencefunctional} are translation invariant, 
they become diagonal in Fourier representation. Our PLD consists in
minimizing the diagonal of $\deco$ for all wave numbers independently. This yields the solution:
\begin{equation}
    \gamma = 2\sqrt{\pairpot \opprod \pairpot} = - 2 \pairpot
\label{eq:PLDstrength}
\end{equation}
because $\pairpot$ is negative definite. Hence, the PLD has led us to 
\begin{equation}
    \deco(\xb,\yb) = \frac{\gamma(\xb,\yb)}{2}=-\pairpot(\xb,\yb)=\frac{G}{\vert\xb-\yb\vert}.
\label{eq:PLDdecoherencefunctional}
\end{equation}
This derivation naturally connects the functional form of the Newtonian interaction potential and of the decoherence kernel, lifting the recent concerns of Gasbarri \textit{et al.} \cite{gasbarri2017} that the noise correlation $\gamma$ should have no connection with the form of the interaction. Here, the unique correlation function $\gamma$,
obtained from the PLD, coincides with the heuristic proposal \eqref{eq:fluctuations} corresponding to the DP model up to a numeric factor. Most importantly,  we have thus obtained a new well grounded derivation of the Di\'osi-Penrose model of gravity-related decoherence and collapse (up to the addition of a gravitational pair potential).

\paragraph{Regularization --}
At that stage, our developments are unfortunately formal because the short distance divergence of the Newtonian potential yields infinite decoherence in the ME \eqref{eq:fullME}, even with our minimal decoherence prescription. A short distance regularization is required in the DP model.
The first option, which was followed in \cite{tilloy2016}, is to mollify $\denso$, \ie ultimately to conjugate both $\gamma$ and $\pairpot$ by the same function $g$ (which can be \eg a Gaussian of width $\sigma$):
\begin{align}
    \deco\rightarrow\deco_\sigma:=&g \opprod \left(\frac{\gamma}{4} + \, \pairpot\opprod g\opprod \left[g\opprod\gamma \opprod g\right]^{-1} \opprod g\opprod \pairpot\right)\opprod g. \nonumber\\
    =&g\opprod \deco \opprod g \label{eq:regulateddecoherence}
\end{align}
The pair potential appearing in the ME \eqref{eq:fullME} is also regularized by this choice:
\begin{equation}
    \pairpot\rightarrow \pairpot_\sigma := g \opprod \pairpot \opprod  g.
\end{equation}
The minimization according to the PLD is clearly independent of the smearing provided we fix the latter before the minimization step, yielding $\gamma=\gamma_\sigma=-2\pairpot_\sigma$ instead of \eqref{eq:PLDstrength}. 
We get the standard regularized DP model with the Newtonian pair potential smeared at short distances.

Insisting on the space-time locality of measurement operators, one may UV regularize the field equations instead of adding a  mollifier $g$ on local operators, say by adding a (quasi-local) biharmonic term $-\sigma^2 \nabla^4\Phi$ on the l.h.s. of the Poisson equation \eqref{eq:poisson}. This is equivalent to taking $\pairpot\rightarrow \pairpot_{\sigma}:=g\opprod \pairpot \opprod g$, this time with $\widetilde{g}(\kb)\propto (1+\sigma^2\, |\kb|^2)^{-1/2}$ in Fourier space (instead of the usual Gaussian). In the end however, this gives the same structural results as before. Indeed, the principle of least decoherence then yields $\gamma = -2 \pairpot_\sigma$ and thus the same decoherence functional as in \eqref{eq:regulateddecoherence}.

Regularization is crucial for such approaches to semi-classical gravity but at the same time the PLD is robust with respect to different choices of regularization procedures. There is a trade-off between the short distance precision of Newton's gravity and decoherence: the smaller the decoherence the larger the distance where
Newton's $1/r$ law breaks down, and vice versa. 
This trade-off makes the model in principle falsifiable for all smearing functions $g$, 
either confirming the pair potential for short distances or confirming the superposition rule
against intrinsic decoherence. This is to be contrasted with standard collapse models \cite{bassi2013} without  our proposed semi-classical gravity, where the parameters yielding slow collapse can only ever be philosophically eliminated \cite{feldmann2012}.

\paragraph{Revisiting LOCC gravity --} We may now revisit the concept of LOCC gravity of Kafri \textit{et al.} in light of our PLD. As we have argued, the PLD breaks locality as the minimal decoherence kernel $\deco$ \eqref{eq:PLDdecoherencefunctional} corresponds to a continuous measurement scheme using spatially entangled detectors. Imposing LOCC would require fixing 
\begin{equation}
\gamma(\xb,\yb) =4\frac{\gamma_\mathrm{CSL}}{m_0^2} \delta(\xb-\yb)
\end{equation}
which would correspond to the (sharp) CSL model in the absence of gravity. For the constant pre-factor we use the CSL notations: $\gamma_\mathrm{CSL}$ is a new parameter and
$m_0$ is the atomic mass unit.
As before, such a choice yields a diverging decoherence without a proper smearing at some scale $\sigma$ so that the concept of LOCC is, strictly speaking, inapplicable for Newtonian semi-classical gravity in the continuum (as was also noted in \cite{kafri2015}). However, slightly relaxing locality up to this distance scale with a function $g$ (which is what the standard CSL model does) yields a decoherence kernel in Fourier space:
\begin{equation}
\widetilde{\deco}(\kb)=\left[\frac{\gamma_\mathrm{CSL}}{m_0^2} +\frac{4m_0^2}{\gamma_\mathrm{CSL}} \cdot \frac{(4\pi G)^2}{|\kb|^4}\right] \widetilde{g}(\kb)^2.
\end{equation}
The first term corresponds to the standard CSL decoherence (here formally coming from the ``$\sigma$-local'' monitoring of the mass density
$\denso$), the second term comes from the noisy gravitational back-reaction  \eqref{eq:potential}. 
Once the smearing factor $ \widetilde{g}(\kb)^2$ is fixed,
the only freedom is in the single parameter $\gamma_\mathrm{CSL}$. There is no unique way to impose a global ``least decoherence'' prescription. However, we may implement some heuristic PLD by requiring that decoherence be minimal for lengths of the order of the smearing length scale $\sigma$. This yields $\gamma_\mathrm{CSL} \sim 8\pi G m_0^2\sigma^2$.

The CSL model defines two new fundamental constants: the smearing parameter $\sigma$  
(often written $r_C$ and called the localization scale) and the collapse rate 
$\lambda_\mathrm{CSL}=\gamma_{\rm CSL}/(4\pi \sigma^2)^{3/2}$. The heuristic
PLD has thus provided the following new relation:
\begin{equation}\label{eq:relationCSL}
\lambda_{\rm CSL}\sim\frac{Gm_0^2}{\sqrt{\pi}\hbar r_C},
\end{equation}
where we have inserted the Planck constant back. This gives a hyperbola in the parameter diagram of CSL that can, again in principle, be experimentally falsified for all values of the smearing scale $\sigma$. Taking the standard localization scale of CSL $ r_C\sim10^{-5}cm$, eq. \eqref{eq:relationCSL} gives $\lambda_{\text{CSL}}\sim 10^{-23}Hz$, which is $7$ orders of magnitude smaller than the standard GRW collapse rate, but still in the ``philosophically satisfactory'' region \cite{feldmann2012}.
\paragraph{Discussion --}

After applying our PLD, all the elbow room is in the short distance regularization. The model we obtain is the least restrictive in terms of decoherence and thus the one that would most convincingly push for some form of quantization of the gravitational field if falsified for all smearing functions. The decoherence predicted by the DP model falsifies it already for regularization distances $\sigma < 10^{-13} cm$ \cite{helou2017}, requiring a cut-off larger than the nucleon size, and thus far larger than the Planck length $l_{\rm P}$. However, the $1/r$-law of gravity below $100\mu m$ has not yet been confirmed (see \cite{Hagedorn2015} for a recent review), hence a blurring $\sigma\ll10^{-2}cm$
is compatible with current data on the Newtonian pair potential. 
Probing gravity at finer scales may be an alternative to noise tests (like \cite{helou2017} and refs. therein)  to falsify its classicality; provided that the PLD is respected by Nature.

Of course, the elephant in the room is still the absence of a relativistic extension of our proposed principle. Continuous measurement and collapse models are notoriously hard to define properly in a relativistic context. Albeit an intense early activity on the subject \cite{diosi1990} in addition to promising recent developments \cite{bedingham2011,pearle2015,tilloy2017}, the formalism that would allow the generalisation of our PLD to fully relativistic contexts does not yet exist. As a result, we just have an uncontrolled low energy approximation and this should be taken into account when considering the fundamental character of the short distance smearing. 

Nonetheless, there are some  indications of robustness of the Newtonian approach from historical relativistic discussions. In a relativistic thought experiment \cite{unruh1984}, Unruh suggested a possible uncertainty relation between the $00$-components of the metric $\mathsf{g}$ and Einstein tensor $\mathsf{G}$:
\begin{equation}\label{eq:Unruh}    
\mathds{E}\left[\delta\bar{\mathsf{g}}_{00}\bar{\mathsf{G}}^{00}\right]\geq\frac{\hbar G}{c^4 VT}.
\end{equation}
where the bar denotes an average on a space-time volume $VT$. In the Newtonian limit, $\delta \mathsf{g}_{00}=2\delta\Phi/c^2$ and $\mathsf{G}^{00}=2\nabla^2\Phi/c^2$, hence $c$ cancels from Unruh's relativistic bound, which reduces to 
\begin{equation}
\mathds{E}\left[(-\overline{\nabla \Phi})^2\right]=\text{const}\times\frac{\hbar G}{VT}.
\label{eq:DL}
\end{equation}
This latter guess was obtained independently and without reference to relativity by Luk\'acs and one of us through a heuristic PLD \cite{diosi1987,diosi1987newtonian,diosi1989minimum}, and
is equivalent to \eqref{eq:fluctuations}.
Similarly in \cite{penrose1996}, Penrose proposed a discussion of space-time uncertainty which, although a priori relativistic, ultimately yielded fluctuations equivalent to \eqref{eq:DL} and
\eqref{eq:fluctuations}. All these lessons suggest that Newtonian semi-classical gravity can be studied autonomously, raising hope that the PLD we discuss is not dramatically modified by  relativistic considerations.

This should not be understood to mean that the general relativistic context is of no interest. On the contrary, collapse models on fixed general gravitational backgrounds have recently proved to be of important theoretical interest \cite{okon2017}. In that case, a mechanism allowing quantum matter to consistently back react on space-time is again the crucial missing piece. Importantly, sourcing gravity with a model analog to ours would give rise to interesting cosmological consequences due to the lack of energy conservation \cite{josset2017}.

\paragraph{Conclusion --}
We have introduced a principle of least decoherence which has allowed us to single out a simple model from a class of consistent theories of Newtonian semi-classical gravity. 
Less demanding than the requirement of LOCC, our principle provides a new well motivated derivation of the DP model. Capitalizing on the standard theory of quantum monitoring and control, we have put on rigorous grounds the heuristic and implicit  principle
the DP model was based upon from its birth.  
Our principle fixes the value of all the parameters but the regularization scale. The latter is the great unknown so far: it determines at what distance gravity becomes blurred and decoherence regularized. Without such a limited precision at short distances (yet $\gg l_{\rm P}$), current semi-classical approaches to gravity are already falsified, unless relativistic effects dramatically modify the present analysis. This calls for an exploration of the short distance behavior of gravity as well as for the development of a theory of hybrid quantum classical dynamics in the relativistic regime.

\bibliography{main}
\bibliographystyle{apsrev4-1}
\end{document}